\documentclass[aps,prb,floatfix,amsmath,amssymb,nofootinbib,twocolumn,reprint]{revtex4}
\usepackage{graphicx}
\usepackage{dcolumn}
\usepackage{bbm}
\usepackage{epstopdf}
\usepackage[bookmarks=true,colorlinks,linkcolor=blue,urlcolor=blue,citecolor=blue]{hyperref}
\usepackage{color}

\usepackage{setspace} 
\newcommand{\beq}{\begin{equation}}
\newcommand{\eeq}{\end{equation}}
\newcommand{\beqn}{\begin{eqnarray}}
\newcommand{\eeqn}{\end{eqnarray}}

\usepackage{amsmath}
\usepackage{amssymb}

\def \q{{\mathbf{q}}}

\def \k{{\mathbf{k}}}

\def \q{{\mathbf{q}}}
\def \p{{\mathbf{p}}}

\begin{document}

\title{Finite temperature scaling close to Ising-nematic quantum critical points in two-dimensional metals}

\author{Matthias Punk}
\affiliation{Physics Department, Arnold Sommerfeld Center for Theoretical Physics, and Center for NanoScience, Ludwig-Maximilians University Munich, Germany}

\date{\today}

\begin{abstract}

We study finite temperature properties of metals close to an Ising-nematic quantum critical point in two spatial dimensions. 
In particular we show that at any finite temperature there is a regime where order parameter fluctuations are characterized by a dynamical critical exponent $z=2$, in contrast to $z=3$ found at zero temperature. 
Our results are based on a simple Eliashberg-type approach, which gives rise to a boson self-energy proportional to $\Omega/\gamma(T)$ at small momenta, where $\gamma(T)$ is the temperature dependent fermion scattering rate. 
These findings might shed some light on recent Monte-Carlo simulations at finite temperature, where results consistent with $z=2$ were found.   

\end{abstract}

\maketitle

\section{Introduction}

Metallic phases that cannot be described within Landau's Fermi liquid framework have been observed in various strongly correlated electron materials, such as cuprate and pnictide superconductors above $T_c$, or heavy fermion compounds.\cite{Hussey, Kasahara, StewartRMP2001} One widely discussed theoretical approach leading to non-Fermi liquid behavior is to couple electrons to gapless fluctuations of an order parameter close to a quantum critical point (QCP).\cite{LohneysenReview} 

In this work we focus on the experimentally relevant example of the so-called Ising-nematic QCP in two spatial dimensions.\cite{FradkinKivelson} Here electrons on a square lattice are coupled to an Ising order parameter, which describes a Pomeranchuk transition where the four-fold rotational symmetry of the Fermi surface is broken down to two-fold rotations. Nematic correlations have been observed in various correlated electron systems, such as underdoped cuprates,\cite{Ando, Khosaka, Hinkov, Daou} and iron-based compounds.\cite{Nandi, Chuang, Chu, Song} Properties of metals close to a nematic QCP, as well as closely related problems such as electrons coupled to a $U(1)$ gauge field, have been extensively discussed in the theory literature.\cite{Lee, AltshulerIoffeMillis, KimFurusakiLee, Oganesyan, Andergassen, Rech, Lawler, DellAnna, Zacharias, SSLee, Metlitski, Maslov2010, Maslov, YamaseMetzner, Bartosch, Hartnoll, LedererSC, Eberlein}

Our main quantity of interest is the nematic susceptibility (i.e.~the retarded propagator of order parameter fluctuations) at small but finite temperatures. Within the Hertz-Millis approach it takes the well known form \cite{DellAnna,Hartnoll,Hertz,Millis}
\begin{equation}
D_R(\k,\Omega)^{-1} = m^2(T) + A \k^2 - i B \cos^2 (2 \varphi_\k) \frac{\Omega}{v_F |\k|} \ ,
\label{eqD0}
\end{equation}
which holds for isotropic systems in the regime $\Omega \ll v_F |\k| \ll \varepsilon_F$, where $\k$ and $\Omega$ are momentum and frequency,  $v_F$ ($\varepsilon_F$) is the Fermi velocity (energy), $m(T)$ is the temperature (T) dependent boson mass (or inverse correlation length) and $A$,$B$ are temperature independent constants. The characteristic Landau damping term $\sim \Omega/|\k|$ arises from the coupling to particle-hole excitations at the Fermi surface. Note that it comes with an angular dependence $\sim \cos^2 2\varphi_\k$, where $\varphi_\k$ is the polar angle of $\k=k (\cos \varphi_\k,\sin \varphi_\k)$, because the coupling between electrons and order parameter fluctuations vanishes by symmetry along the nodal directions $k_x=\pm k_y$. At the QCP where $m(0)$ vanishes, the susceptibility obeys scaling with a dynamical critical exponent $z=3$, i.e.~it is invariant under rescaling $\k' = b \k$ and $\Omega' = b^{z} \Omega$. 

At zero temperature Eq.~\eqref{eqD0} also holds beyond the random phase approximation (RPA) in higher order perturbation theory, as well as within a self-consistent Eliashberg-type approximation, where bare propagators are replaced with full propagators.\cite{Polchinski} 
More recent works focusing on the critical properties at zero temperature realized that the Hertz-Millis RPA approach has  conceptual problems, however. While it was previously believed that the RPA can be justified in a large $N_F$ limit,\cite{AltshulerIoffeMillis, KimFurusakiLee} where $N_F$ is the number of fermion flavors, it has been realized that such large $N_F$ expansions break down due to intricate quasi one-dimensional scattering processes in certain subsets of Feynman diagrams.\cite{SSLee, Metlitski} Subsequently several approximation schemes have been developed where controlled expansions in a small parameter can be performed, such as a combination of small $\varepsilon=z-2 $ and $1/N_F$ with $N_F (z-2)$ fixed,\cite{Mross} an expansion in a large number of boson flavors $N_B$,\cite{Raghu} or performing an epsilon expansion in the co-dimension of the Fermi surface.\cite{Dalidovich}

Despite these problems, the structure of Eq.~\eqref{eqD0} is compatible with renormalization group results where bosons and fermions are treated on equal footing. The dynamical critical exponent $z=3$ remains unchanged up to three loops,\cite{Metlitski} even though more recent four-loop results indicate that anomalous scaling appears at higher loop order.\cite{Holder} It is important to note that these RG approaches typically deal with the zero temperature problem. Finite temperature results are usually inferred by assuming $\omega/T$ scaling in the vicinity of the QCP. This assumption can be potentially dangerous, however. For example, it has been shown that the electron scattering rate at the Fermi surface is dominated by contributions from \emph{classical} fluctuations at low temperatures, which do not obey $\omega/T$ scaling.\cite{DellAnna}

Substantial progress towards a numerical solution was made by Schattner \emph{et al.},\cite{Schattner} who realized that the Ising-nematic problem is amenable to unbiased Monte Carlo simulations avoiding the infamous fermion sign problem. Surprisingly, the finite temperature form of the nematic susceptibility found in this work is consistent with a dynamical critical exponent $z=2$, rather than $z=3$ obtained in previous field theoretical approaches at $T=0$. It remains to be seen if the temperatures in the numerical simulations are low enough to probe the scaling regime of the QCP, or if something is missing in the field theoretical approaches.

Here we investigate finite temperature properties of the nematic susceptibility in the quantum critical regime above the QCP. We focus on the important interplay between bosonic and fermonic excitations at finite temperature using an Eliashberg-type approach, where the boson and fermion self-energies are computed in a self-consistent one-loop approximation. At finite temperature the Eliashberg approach is shown to give qualitatively different results compared to RPA, in contrast to the zero temperature case.
In particular, the nematic susceptibility takes the form
\begin{equation}
D_R(\k,\Omega)^{-1} = m^2(T) +A \k^2 -i B \frac{\Omega}{\gamma(T)}
\label{eqDs}
\end{equation}
for $v_F |\k| \ll \gamma(T)$, where $\gamma(T) \sim \sqrt{T / |\log T|}$ is the temperature dependent electron scattering rate and $v_F$ is the Fermi velocity. This result suggests that at any finite temperature there is always an energy window where the order parameter fluctuations are characterized by a dynamical critical exponent $z=2$, rather than $z=3$ at zero temperature \cite{footnote}. In the zero temperature limit where $\gamma(T)$ vanishes we recover Eq.~\eqref{eqD0}. 

A potential shortcoming of the result in Eq.~\eqref{eqDs} is that the frequency dependent term doesn't vanish in the limit $\k \to 0$. This is a particularly severe problem for the closely related problem of electrons at a ferromagnetic QCP, where the order parameter is conserved. Indeed, the Eliashberg approach is an uncontrolled approximation where potentially important vertex corrections are neglected, which can lead to violations of Ward identities. 
However, in the important limit $\Omega \ll v_F |\k|$ the boson velocity is small compared to the Fermi velocity and standard arguments in analogy to Migdal's theorem should apply, which ensures the smallness of vertex corrections, at least in the zero temperature limit.\cite{Polchinski,AltshulerIoffeMillis,KimFurusakiLee,Rech} 

At finite temperature the situation is different, because classical (frequency independent) fluctuations - which are not present at zero temperature - dominate the vertex correction. Consequently, our problem is seemingly similar to the disordered electron gas, where vertex corrections are large and the vertex develops a diffuson pole. It is important to emphasize, however, that in the quantum critical regime both classical and quantum fluctuations are equally important. In fact, we argue that the vertex does not develop a diffuson pole if quantum fluctuations are taken into account and the results of the Eliashberg approximation remain qualitatively valid.

The remainder of this paper is outlined as follows. In Sec.~\ref{sec2} we introduce the model of electrons coupled to an Ising nematic order parameter and introduce the Eliashberg approach used subsequently. Sec.~\ref{sec3} contains analytical results for the electron and boson self-energies, whereas a numerical results are presented in Sec.~\ref{sec4}. Finally, vertex corrections at finite temperature are discussed in Sec.~\ref{sec5}. Conclusions are presented in Sec.~\ref{sec6}.

\section{Model and Methods}
\label{sec2}

We start from a model of spin-$1/2$ electrons on the square lattice coupled to an Ising nematic order parameter described by the euclidean action (spin index suppressed)
 \begin{eqnarray}
S &=&\sum_{\k,\omega_n} \bar{c}_{\k,\omega_n} (-i \omega_n +\xi_\k) c_{\k,\omega_n} \notag \\
&& + \frac{1}{2} \sum_{\k,\Omega_n}  \chi^{-1}_\k \, \phi_{\k,\Omega_n} \phi_{-\k,-\Omega_n} \notag \\
&& + \frac{\lambda}{\sqrt{\beta V}} \sum_{\substack{\k,\omega_n \\ \q,\Omega_n}} d_\q \, \phi_{\k,\Omega_n}  \bar{c}_{\q+\k/2,\omega_n+\Omega_n}  c_{\q-\k/2,\omega_n} \, . \quad \quad
\label{eq:model}
\end{eqnarray}
Here the fermionic fields $c_{\k,\omega_n}$ describe electrons with momentum $\k$ and Matsubara frequency $\omega_n$, where $\xi_\k =\varepsilon_\k - \mu$ is the electron dispersion measured from the chemical potential, and the scalar field $\phi_{\k,\Omega_n}$ represents the Ising-nematic order parameter. The static nematic susceptibility $\chi_\k = 1/\big[m^{2} + 2 A (2-\cos k_x -\cos k_y)\big]$, i.e.~the bare propagator of the $\phi_k$ field, is chosen to be maximal at $\k=0$ and consistent with square lattice symmetry and $\lambda$ parametrizes the coupling strength. $\beta=1/T$ is the inverse temperature, $V$ the volume (we use natural units $\hbar=k_B=1$ and set the lattice constant to unity throughout) and 
\begin{equation}
d_\q = \cos q_x - \cos q_y
\end{equation}
is the nematic d-wave form-factor.
Together with the Ising symmetry $\phi \to -\phi$ the action \eqref{eq:model} is symmetric under $90^\circ$ rotations. The characteristic d-wave form-factor $d_\q$ in the interaction term leads to a breaking of the four-fold rotation symmetry in the ordered phase with $\langle \phi_0 \rangle \neq0$. 

The electron propagator and the nematic fluctuation propagator are given by 
\begin{eqnarray}
G(\k,i \omega_n) &=& \frac{1}{i \omega_n - \xi_\k-\Sigma(\k,i\omega_n)} \\
D(\k,i \Omega_n) &=& \frac{1}{\chi_\k^{-1} - \Pi(\k,i \Omega_n)} \ .
\end{eqnarray}
In the following we compute the fermionic and bosonic self-energies $\Sigma(k)$ and $\Pi(k)$ using a self-consistent one-loop approximation, neglecting vertex corrections. The coupled Eliashberg equations for the self-energies take the form
\begin{eqnarray}
\Sigma(k) &=& \frac{\lambda^2}{\beta V} \sum_{q} G(k-q) D(q) \, d^2_{\k-\q/2} \label{eq:sigma} \\
\Pi(k) &=& - 2 \frac{\lambda^2}{\beta V} \sum_{q} G(k+q) G(q) \, d^2_{\q+\k/2}  \label{eq:pi} \ ,
\end{eqnarray}
where we use the shorthand notation $k=(i \omega_n,\k)$ etc., and the factor of two is for spin. After analytic continuation $i \omega_n \to \omega + i 0^+$ to real frequencies the equations for the imaginary parts of the retarded self-energies take the form
\begin{eqnarray}
\text{Im} \Sigma_R(\mathbf{k},\omega) \! &=& \! \lambda^2 \int_\q \int \frac{dz}{\pi} \big[ n_B(z) + n_F(z-\omega)\big] \, d^2_{\k-\q/2}\notag \\ && \! \times \,\text{Im} G_R(\k-\q,\omega-z) \, \text{Im} D_R (\q,z)  \label{eq1} \\
\text{Im} \Pi_R(\k,\Omega) \! &=& \! 2 \lambda^2 \int_\q \int \frac{dz}{\pi} \big[ n_F(z) - n_F(z+\Omega)\big] \, d^2_{\q+\k/2} \notag \\ 
&& \!  \times \,\text{Im} G_R(\k+\q,\Omega+z) \, \text{Im} G_R (\q,z) \, ,  \label{eq2}
\end{eqnarray}
where $n_B(z)$ and $n_F(z)$ are the Bose-Einstein and Fermi-Dirac distribution functions and $\int_\q \equiv \int_\text{BZ} \frac{d^2q}{4 \pi^2}$ denotes a momentum integral over the first Brillouin zone.

\section{Analytical Results}
\label{sec3}

We start by deriving analytic results for the simpler case of a circular Fermi surface, $\xi_\q = (\q^2-k_F^2)/2m_e$, with $k_F$ the Fermi momentum and $m_e$ the electron mass. Our aim is to show that at nonzero temperatures electron excitations have a finite lifetime due to the interaction with thermally excited bosons, which in turn changes the momentum dependence of the boson self-energy drastically compared to $T=0$.

The inverse electron lifetime, i.e.~the imaginary part of the electron self-energy at the Fermi energy $\gamma_{\mathbf{k}_F} \equiv -\text{Im}\Sigma_R(\mathbf{k}_F,0)$ follows from Eq.~\eqref{eq1}. Observing that the integral is dominated by contributions from small frequencies at low temperatures, we expand in small $z$ and obtain
\begin{eqnarray}
\gamma_{\mathbf{k}_F} &\simeq& -\lambda^2 \int_\q \int \frac{dz}{\pi} \, \frac{1}{\beta z} \frac{\text{Im} \Pi_R (\q,z)}{(m^{2}+A \q^2)^2+\text{Im} \Pi_R^2 (\q,z)} \notag \\
&&  \times \, \text{Im} G_R(\k_F-\q,0) \, d^2_{\k_F-\q/2} \ .
\label{eqgamma0}
\end{eqnarray}
Note that we neglect the real parts of all self-energies in analytic computations for simplicity. Quite generically, the imaginary part of the boson self-energy at small frequencies takes the form
\begin{equation}
\text{Im} \Pi_R (\q,|z|\ll1) = z \, \mathcal{P}(\q) \ ,
\label{eqasymp1}
\end{equation}
with an as yet unknown function $\mathcal{P}(\q)$, which has the well known Landau damping form $\mathcal{P}(\q)\sim 1/|\q|$ at zero temperature. Using \eqref{eqasymp1} we can perform the frequency integral in \eqref{eqgamma0} straightforwardly and obtain
\begin{equation}
\gamma_{\mathbf{k}_F} \simeq \frac{\lambda^2}{\beta} \int_\q  \frac{1}{m^{2}+A \q^2} \, \frac{\gamma_{\k_F-\q}}{\xi_{\k_F-\q}^2+\gamma_{\k_F-\q}^2} \, d^2_{\k_F-\q/2} \ .
\end{equation}
Note that $\mathcal{P}(\q)$ drops out of this expression for $\gamma_{\k_F}$, i.e.~the momentum dependence of the boson self-energy doesn't play any role for the electron lifetime.
At low enough temperatures, where $\gamma_{\k_F}$ is much smaller than the Fermi energy, the dominant contribution to the integral above comes from momenta close to the Fermi surface, i.e. from small $\q$, and can be approximated as
\begin{equation}
\gamma_{\k_F} \simeq \frac{\lambda^2}{4 \pi^2 \beta} \int_0^\infty \! dq \int_0^{2 \pi} \! d\theta \frac{q}{m^{2}+A q^2} \frac{\gamma_{\k_F} \,  d_{\k_F}^2}{(v_F q \cos \theta)^2 + \gamma_{\k_F}^2}
\label{eqgamma}
\end{equation}
where we've expanded in small momenta $q$ and $v_F=k_F/m_e$ is the Fermi velocity. Evaluating the integrals we finally obtain
\begin{equation}
\gamma_{\mathbf{k}_F} \simeq  \frac{\lambda^2}{4 v_F A^{1/2}} \, \frac{T}{m(T)}  \, d_{\k_F}^2  \quad \text{for} \quad  \frac{\lambda^2 T}{v_F^2 m^2(T) } \ll 1 \ . 
\label{gammakf}
\end{equation}
This reproduces earlier results by Dell'Anna and Metzner.\cite{DellAnna}
Note that $\gamma_{\k_F}$ has a momentum dependence along the Fermi surface due to the d-wave form-factor $d_{\k_F}$. In particular $\gamma_{\k_F} = 0$ in the nodal directions $k_x=\pm k_y$, since electrons do not interact along these momenta.
In the quantum critical regime the temperature dependence of the boson mass is expected to take the form $m(T) \sim \sqrt{T \, | \! \log T|}$.\cite{Millis,Hartnoll} Consequently, the electron scattering rate scales as $\gamma(T) \sim \sqrt{T/| \! \log T|}$ at low temperatures. 

\begin{figure}
\begin{center}
\includegraphics[width=0.95 \columnwidth]{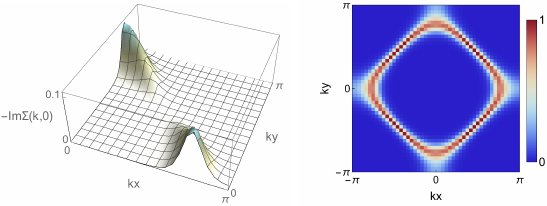}
\caption{(color online) Left: $-\text{Im} \Sigma_R (\k,0)$ at the Fermi energy as a function of momenta in one quadrant of the Brillouin zone. Right: imaginary time Green's function $2 G(\k,\tau=\beta/2)$. Both results were obtained from a numerical solution of Eqs.~\eqref{eq1} and \eqref{eq2} at a temperature $T/t=0.1$, the remaining parameters are specified in the main text.}
\label{fig1}
\end{center}
\end{figure}

It is important to realize that the we only considered the pole contribution of the Bose distribution in evaluating the electron scattering rate in Eq.~\eqref{eqgamma0}. This amounts to taking only the interaction of electrons with classical (frequency independent) fluctuations of the boson mode into account. The interaction with quantum fluctuations gives rise to a subleading $\sim T^{2/3}$ dependence, which we omit in the following. 

We now move on to the boson self-energy. For small external frequencies and momenta $\Omega \ll v_F |\k| \ll \varepsilon_F$ we expand the integrand in \eqref{eq2} and at low enough temperatures we obtain
\begin{equation}
\text{Im} \Pi_R(\k,\Omega) \simeq 2 \lambda^2 \Omega \int_\q   \delta(\xi_\q) \frac{\gamma_{\q} \, d_{\q}^2 }{(\q \cdot \k/m_e)^2+\gamma_{\q}^2} \ , \
\end{equation}
where we assumed that $\gamma_\q \ll \varepsilon_F$ is small enough to replace one Lorentzian with a delta function, which pins the absolute value of $\q$ to the Fermi momentum $k_F$. Making the angular dependence of $\gamma_{\q_F}$ explicit by writing $\gamma_{\q_F} = \gamma \, d^2_{\q_F}$ and using the simplified d-wave form factor $d_{\q_F} = \cos 2 \theta$, where $\theta$ is the polar angle of $\q_F =k_F (\cos \theta, \sin \theta)$, we arrive at
\begin{equation}
\text{Im} \Pi_R(\k,\Omega) \simeq \frac{\lambda^2 m_e}{\pi} \frac{\Omega}{\gamma(T)} \, \tilde{\Pi} \! \left(\frac{v_F |\k|}{\gamma(T)},\varphi_\k \right) \ ,
\label{mainresult}
\end{equation}
where $\varphi_\k$ is the polar angle of $\k$ and the scaling function $\tilde{\Pi}$ takes the form
\begin{equation}
\tilde{\Pi}(x,\varphi) = \int_0^{2 \pi} \frac{d\theta}{2 \pi} \, \frac{\cos^4 2\theta}{x^2 \cos^2(\theta-\varphi)+ \cos^4 2\theta} \ .
\label{scalingfunc}
\end{equation}
It has the limiting forms
\begin{eqnarray}
\tilde{\Pi}(x \to 0,\varphi) &=& 1 \\
\tilde{\Pi}(x \gg 1,\varphi) &=& \frac{\cos^2 2 \varphi}{x} \ . 
\end{eqnarray}
At zero temperature, where $\gamma$ vanishes, we thus recover the standard Landau damping form.

\begin{figure}
\begin{center}
\includegraphics[width=0.85 \columnwidth]{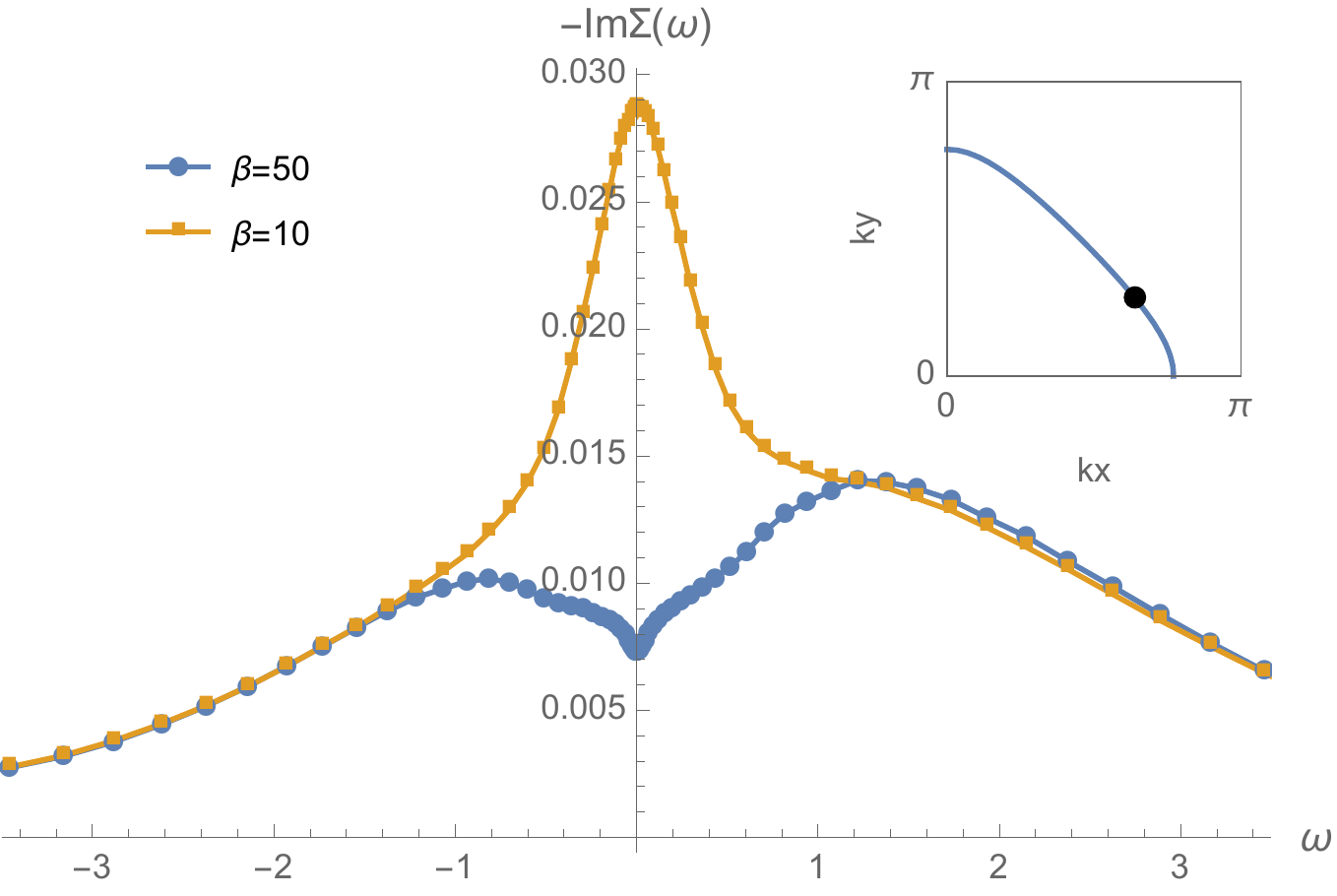}
\caption{(color online) Imaginary part of the electron self-energy $-\text{Im} \Sigma_R (\k,\omega)$ as function of frequency $\omega$ at momentum $\k=(k_F,0)$ on the Fermi surface (indicated by the point in the inset), shown for two temperatures $T/t=0.02 \ (\beta=50)$ and $T/t=0.1 \ (\beta=10)$ (other parameters specified in the main text).}
\label{fig2}
\end{center}
\end{figure}

Our main result in Eq.~\eqref{mainresult} indicates that in the momentum regime $\gamma(T)/v_F \gg |\k| \gg m(T)/A^{1/2} $ the bosonic excitations at finite temperature are characterized by a dynamical critical exponent $z=2$. 
Using $m^2(T) \sim T | \log T |$ it's easy to see from Eq.~\eqref{eqgamma} that this regime only appears at intermediate temperatures $T \gg \varepsilon_F \exp(-\lambda^2 / 2 \pi v_F^2)$. Note that for large enough couplings $\lambda$ this intermediate $z=2$ regime can be found at arbitrary low temperatures.
Interestingly, the fact that this regime doesn't extend asymptotically to $T=0$ relies crucially on the $\log$ correction to the boson mass $m(T)$. Indeed, for $m^2(T) \sim T$ the momentum regime with $z=2$ scaling would extend asymptotically to zero temperature.

\section{Numerics}
\label{sec4}

\begin{figure}
\begin{center}
\includegraphics[width=0.9 \columnwidth]{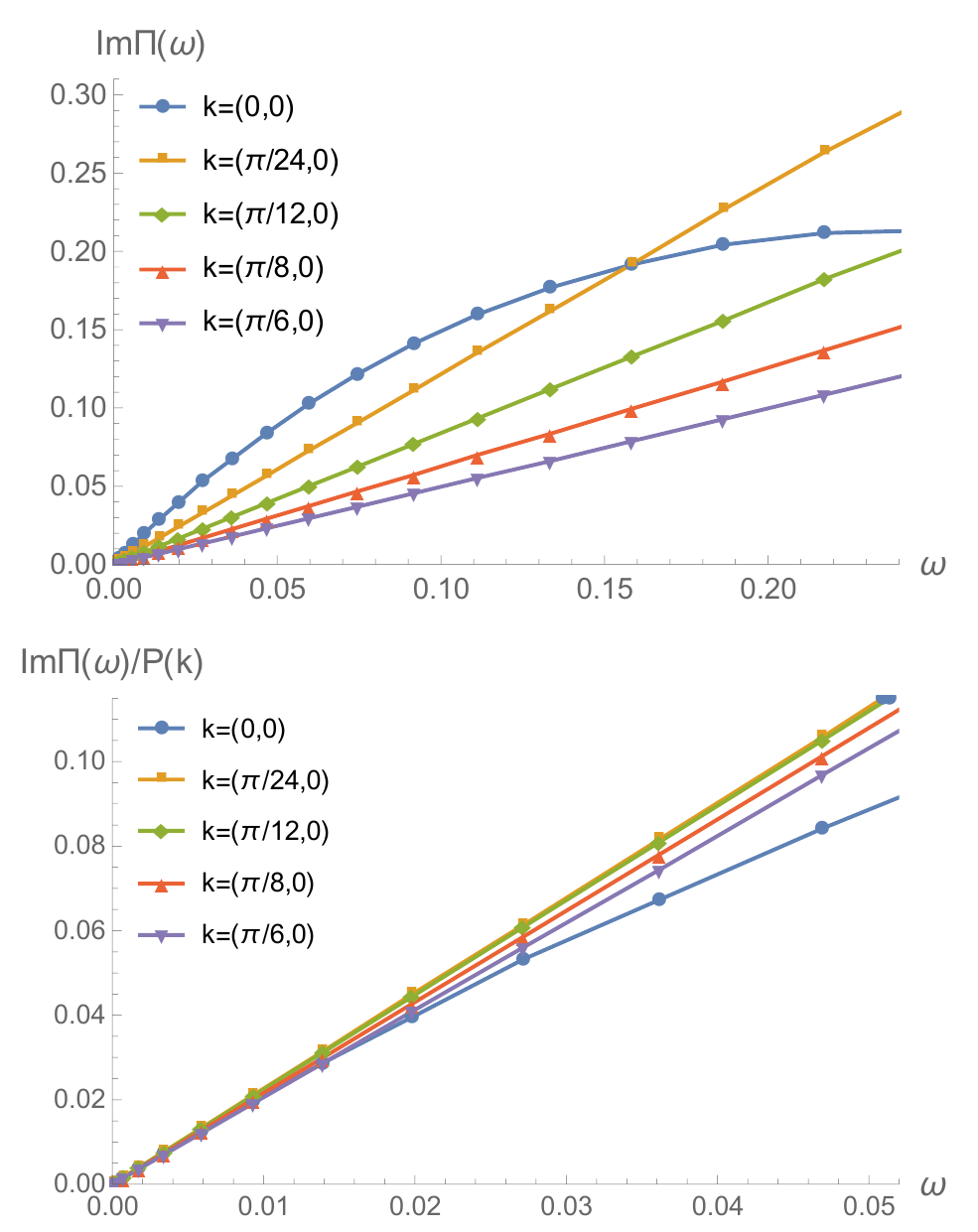}
\caption{(color online) Upper panel: imaginary part of the boson self-energy $\text{Im} \Pi_R (\k,\omega)$ as a function of frequency $\omega$ for different momenta $\k$ at temperature $T/t=0.2$. Lower panel: same data scaled with the scaling function in Eq.~\eqref{scalingfunc}, with $P(k)=\tilde{\Pi}(k/0.25,0)$, showing data collapse at small frequencies (other parameters specified in the main text).}
\label{fig3}
\end{center}
\end{figure}

In order to support our analytical considerations, we present results from a full numerical solution of the Eliashberg equations \eqref{eq1} and \eqref{eq2} in the following. We use the same strategy as in Ref.~[\onlinecite{Punk}] and solve the equations on a discretized grid of $49\times49\times101$ points in momentum and frequency space, with a non-linear discretization in frequency space to obtain a better resolution at small frequencies. We choose a nearest neighbor tight binding dispersion
$\xi_\q = -2 t (\cos q_x + \cos q_y) - \mu$, measure energies in units of $t=1$, set the renormalized chemical potential $\mu -\text{Re} \Sigma_R(\k_{F}^\text{node},0)=-0.5$ and the coupling constant to $\lambda=0.7$. Furthermore we do not compute the boson mass $m(T)$ selfconsistently, but fix the renormalized gap $\tilde{m}^2 = m^2 - \text{Re} \Pi_R(\mathbf{0},0)$ at $\tilde{m}^2=0.01,0.02$ and $0.025$ for inverse temperatures $\beta=50,10$ and $5$, for which data is shown here.

In Fig.~\ref{fig1} we show the electron scattering rate at the Fermi energy $\gamma_\k = - \text{Im} \Sigma(\k,0)$ as function of momenta. Note that $\gamma_\k$ is maximal along the Fermi surface and has the characteristic angular dependence expected from Eq.~\eqref{gammakf}. The right panel shows two times the imaginary time electron Green's function at imaginary time $\tau=\beta/2$, which reduces to the electronic quasiparticle residue at zero temperature. Indeed, $2 G(\k,\tau=\beta/2)= \int d\omega A(\k,\omega)/\cosh(\beta \omega/2)$, where $A(\k,\omega) = -\pi \text{Im} G_R(\k,\omega)$ is the electron spectral function. This quantity can be compared directly to the Monte Carlo results in Ref.~[\onlinecite{Schattner}] and agrees nicely.

Fig.~\ref{fig2} shows the imaginary part of the electron self-energy for one point on the the Fermi surface as function of frequency for two different temperatures. Note that at zero temperature it should scale as $\text{Im} \Sigma(\k_F,\omega) \sim \omega^{2/3}$ at small frequencies. Indications of this scaling behavior can be seen already at $T/t=0.02$, but it is cut off by the finite electron scattering rate at the Fermi energy $\omega=0$.
Lastly, we plot the imaginary part of the boson self-energy $\text{Im} \Pi_R(\k,\Omega)$ at inverse temperature $\beta=5$ as function of frequency for various momenta in Fig.~\ref{fig3}. The lower panel displays the rescaled data using the scaling function from Eq.~\eqref{scalingfunc} with $\gamma/v_F = 0.25$, showing scaling collapse at small frequencies.

\section{Vertex Corrections}
\label{sec5}

\begin{figure}
\begin{center}
\includegraphics[width=0.6 \columnwidth]{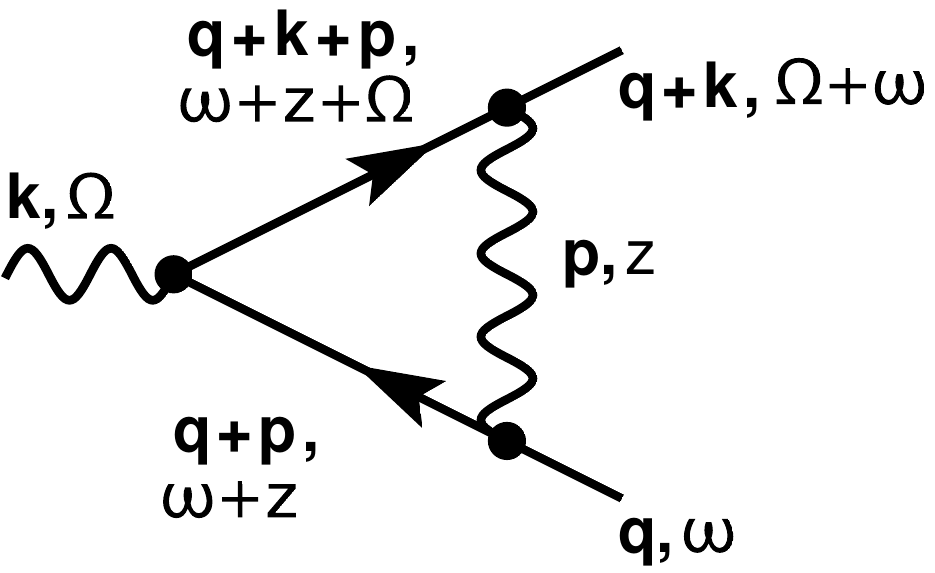}
\caption{Diagrammatic representation of the leading one loop vertex correction $\Gamma^{(1)}(\k,\Omega;\q,\omega)$.}
\label{fig4}
\end{center}
\end{figure}

In order to assess the validity of the Eliashberg approximation we compute the leading one loop vertex corrrection shown in Fig.~\ref{fig4} at finite temperature, using the approximate form of the propagators obtained in Sec.~\ref{sec3}, in particular 
\begin{eqnarray}
G^{-1}(\k,i\omega_n) &=& i\omega_n - \xi_\k + i \gamma_{\k_F} \text{sgn}(\omega_n) \\
D^{-1}(\k,i \Omega_n) &=& m^2 + A \k^2 + B \frac{|\Omega_n|}{\gamma} \label{propE} \ . \\
\notag
\end{eqnarray}
The one loop vertex correction $\Gamma^{(1)}(\k,i \Omega_n;\q, i \omega_n)$ takes the form 
\begin{eqnarray}
\Gamma^{(1)}(k;q) &=& \frac{\lambda^3}{\beta V} \sum_p D(p) G(q+p) G(q+p+k) \notag \\
&& \ \  \times d_{\q+\p/2} d_{\q+\p+\k/2} d_{\q+\p/2+\k}
\label{vertex0}
\end{eqnarray}
where we use the shorthand notation $k=(i \Omega_n,\k)$, $q=(i \omega_n,\q)$, etc.
Since the important scattering processes involve electrons in the vicinity of the Fermi surface with small momentum transfer, we only consider the vertex at vanishing external boson momentum $\k=0$ and external electron momentum on the Fermi surface $\q=\k_F$, as well as take the limit of a vanishing incoming electron frequency $\omega \to 0$ in the following. Furthermore, realizing that the momentum integral in Eq.~\eqref{vertex0} is dominated by small momenta, we expand the d-wave form factors in small $\p$ and only retain the leading order term $\sim d_{\k_F}^3$.

For the following discussion it is convenient to split the vertex correction into a classical and a quantum part $\Gamma = \Gamma_\text{cl}+\Gamma_\text{qu}$. The classical part comes from the zero frequency term in the Matsubara sum in Eq.~\eqref{vertex0}, whereas the summands with non-zero frequencies describe contributions from quantum fluctuations of the boson mode and constitute the quantum part of the vertex.
While classical fluctuations don't exist at zero temperature (where the Matsubara sum becomes an integral and the zero frequency term is a set of measure zero), they are actually dominant at non-zero temperatures, as we'll show in the following.

The computation of the classical part of the vertex correction is analogous to the problem of the disordered electron gas and the largest contribution comes from the term where one electron propagator is retarded and the other one is advanced. Using the approximations mentioned above and performing the analytic continuation $i \omega_n \to \omega + i 0^-$ and $i (\omega_n+\Omega_n) \to \omega+\Omega + i 0^+$ as well as taking the limit $\omega \to 0$,
the classical part of the vertex correction takes the form
\begin{widetext}
\begin{eqnarray}
\Gamma^{(1)}_\text{cl}(0,\Omega;\k_F,0) &\simeq& \frac{\lambda^3 d_{\k_F}^3}{\beta} \int \frac{d^2p}{4 \pi^2} D(\p,0) G_A(\k_F+\p,0) G_R(\k_F+\p, \Omega) \notag \\
&\simeq&  \frac{\lambda^3 d_{\k_F}^3}{4 \pi^2 \beta}  \int_0^\infty  dp \, p \int_0^{2 \pi} d\theta \, \frac{1}{m^2+a p^2} \, \frac{1}{v_F^2 p^2 \cos^2 \theta + \gamma_{\k_F}^2} \left(1 + \frac{ \Omega}{v_F p \cos \theta -i \gamma_{\k_F} }+ \dots \right) \notag \\
&=& \lambda \, d_{\k_F} \left( 1 + \frac{i \Omega}{\gamma_{\k_F}} + \dots \right) \ \ \ \ \ \text{for} \ \ \ \  \frac{\lambda^2 T}{v_F^2 m^2} \ll 1 \ ,
\label{vertex1}
\end{eqnarray}
\end{widetext}
where the dots denote higher order terms in $\Omega$. 
The classical contribution to the leading vertex correction at zero external frequency is temperature independent and equal to the bare vertex $\lambda \, d_{\k_F}$. Consequently it is not negligible and vertex corrections need to be resummed to all orders in perturbation theory. Performing a ladder resummation of the classical vertex correction and neglecting the quantum contribution, the vertex would be given by
\begin{equation}
\Gamma_\text{cl}^\text{ladder} (0,\Omega;\k_F,0)= \frac{\lambda \, d_{\k_F}}{1- \frac{\Gamma^{(1)}_\text{cl}(0,\Omega;\k_F,0)}{\lambda  d_{\k_F}}} = \frac{\lambda \, d_{\k_F}}{-i \Omega/\gamma_{\k_F} } \ .
\label{vertCL}
\end{equation}
Note that the classical vertex develops the well known diffuson pole, familiar from the theory of the disordered electron gas.

The presence of a diffuson pole would invalidate the Eliashberg approximation and change the scaling properties of the boson self-energy drastically. It is important to realize, however, that the problem of the Ising-nematic quantum critical metal at finite temperature differs from the disordered electron gas in one crucial aspect: in the quantum critical regime at finite temperature both classical and quantum fluctuations are equally important and it is not permissible to neglect the contribution from quantum fluctuations, even though the classical fluctuations dominate.
In order to estimate the quantum correction to the vertex, we simply compute the one loop diagram in Fig.~\ref{fig4} at zero temperature using the approximate zero temperature form of the propagators (i.e.~set $\gamma_{\k_F}=0$, $m=0$ and use $|\Omega|/v_F |\k|$ instead of $|\Omega|/\gamma$ in Eq.~\eqref{propE}). At vanishing external frequencies we obtain
\begin{eqnarray}
\Gamma^{(1)}_\text{qu} &\equiv& \Gamma^{(1)}_\text{qu} (0,0,\k_F,0) \notag \\
&\simeq& \lambda^3 d_{\k_F}^3 \int \frac{d^2 p}{4 \pi^2} \int \frac{dz}{2\pi} G^2(\k_F+\p,i z) D(\p,i z) \notag \\
&=& - \frac{\lambda^3 d_{\k_F}^3}{ v_F \sqrt{A B}} \, 0.0674\dots
\end{eqnarray} 
Note that the quantum correction comes with a negative sign. Again performing a ladder resummation of the vertex correction taking into account both classical and quantum contributions, the vertex now takes the form
\begin{equation}
\Gamma^\text{ladder}(0,\Omega;\k_F,0) \simeq \frac{\lambda \, d_{\k_F}}{-\Gamma^{(1)}_\text{qu}(0,0;\k_F,0) - \frac{i \Omega}{ \gamma_{\k_F}}} 
\end{equation}
The crucial difference to Eq.~\eqref{vertCL} is that the quantum correction gaps out the diffuson pole of the classical vertex. More importantly, using this form of the vertex to compute the boson self energy beyond the Eliashberg approximation doesn't alter the scaling properties of the imaginary part derived in Sec.~\ref{sec3} as long as $\Omega/\gamma_{\k_F}$ is small compared to the quantum vertex correction. Consequently we expect that the Eliashberg approximation gives qualitatively valid results at finite temperature and small frequencies.

\section{Conclusions}
\label{sec6}

In this work we considered two dimensional metals in the vicinity of an Ising-nematic quantum critical point and discussed properties of the nematic susceptibility at finite temperature based on an Elisahberg-type approach.
Our results have some similarities to the Monte Carlo results by Schattner \emph{et al.},\cite{Schattner} who find a nematic susceptibility consistent with $z=2$ scaling in the quantum critical regime and no angular dependence on momentum. However, in order for our results to be consistent with their data, the electron scattering rate has to be of order $\varepsilon_F$ even at the the lowest temperatures, otherwise the crossover to $\Omega/|\q|$ behavior should be visible in the Monte Carlo data at large momenta $\sim k_F$. In our numerical solution of the Eliashberg equations shown in Fig.~\ref{fig3} the scattering rate is always substantially smaller than $\varepsilon_F$, even at relatively high temperatures, and the crossover to the standard Landau damping form is always visible. It thus remains to be seen if the phenomenology discussed here is indeed responsible for the behavior observed in the Monte Carlo data.

\acknowledgements

We are grateful for very helpful discussions with D.~Chowdhury, A.~Chubukov, and E.~Berg. Furthermore we'd like to thank an anonymous referee for drawing our attention to the analogy between classical vertex corrections and the disordered electron gas. This work is supported by the Nano-Initiative Munich (NIM).


\begin{thebibliography}{999}
\bibitem{footnote}
In the large $N_B$ expansion of Ref.~\cite{Raghu} an intermediate regime with $z=2$ appears as well, which might be related.
\end{thebibliography}
\end{document}